\documentclass[%
reprint,                                         
 amsmath,amssymb,
 aps,    prc
]{revtex4-2}

\usepackage{graphicx}
\usepackage{dcolumn}
\usepackage{bm}

\usepackage{multirow}

\usepackage{rotating}

\begin{document}

\title{Fine-tuning of the $\bar{K}NN$ and $\bar{K}\bar{K}N$ quasi-bound state calculations}

\author{N.V. Shevchenko}%
 \email{shevchenko@ujf.cas.cz}
\affiliation{%
 Nuclear Physics Institute of the Czech Academy of Sciences, 25068 \v{R}e\v{z}, Czech Republic
}%

\date{\today}

\begin{abstract}
Fine-tuning of the binding energies and widths of the quasi-bound states in three-body
systems consisting of antikaon(s) and nucleon(s) was performed. Dynamically exact
three-body Faddeev-type AGS equations with three coupled particle channels were solved
for the description of the $\bar{K}NN$ and $\bar{K} \bar{K} N$ systems in different spin states.
New models of the antikaon-nucleon and pion-nucleon interactions were constructed, and
together with our best version of the hyperon-nucleon potential, were used as input. The characteristics
of the quasi-bound $K^- pp$ state calculated with our new one-pole $\bar{K}N - \pi \Sigma - \pi \Lambda$
potential reproduces the experimental data from the E15 J-PARC experiment.
\end{abstract}



\maketitle

\section{Introduction}
\label{intro_sect}

Exotic systems consisting of mesons and nucleons can provide additional information on the meson-nucleon
interaction, which is hard to study in scattering experiments.
The existence of a quasi-bound state, a bound state with non-zero width, was predicted for systems consisting
of antikaon(s) and nucleon(s). Many theoretical efforts were devoted to the study of the lightest possible $K^- pp$
system, different methods and inputs were used.  Due to this, the predicted binding energies and
widths differ from each other, but all theoretical calculations agree that the quasi-bound $K^- pp$ state can exist.
Several experiments reported evidence of the state observation, and the most recent E15 experiment at J-PARC \cite{E15_JPARC1,E15_JPARC2} reported the first clear signal of the $\bar{K}NN$ quasi-bound state.
While the measured binding energy of the state $B_{\bar{K}NN} \approx 40$ MeV is comparable with some
of the theoretical predictions, the experimental width $\Gamma_{\bar{K}NN} \approx 100$ MeV is much larger than
all of them.

Our studies of the $K^- pp$ system~\cite{my_review} were performed using dynamically exact three-body
Faddev-type Alt-Grassberger-Sandhas (AGS) equations \cite{AGS3} with coupled $\bar{K}NN - \pi \Sigma N$ channels
with different input. Namely three-body Faddeev-type equations in momentum representation are perfect
for the study of few-body systems consisting of antikaon(s) and nucleon(s).
The predicted binding energies of the $K^- pp$ state \cite{Kd_Kpp_last} calculated using two of three our antikaon-nucleon
potentials are close to the experimental values \cite{E15_JPARC1,E15_JPARC2}. However, the predicted widths
are much smaller than the experimental one measured by the E15 experiment.

We tried to resolve the question of whether it is possible to obtain theoretical results closer to
the experimental ones. In our previous calculations~\cite{my_review}, we demonstrated that  the antikaon-nucleon interaction
plays the main role in the description of the $K^- pp$ and other few-body systems consisting of antikaon(s) and nucleon(s);
on the contrary, dependence on the nucleon-nucleon potential is weak.
Recently \cite{FineTune_FBS}, we additionally studied the dependence of the characteristics of the quasi-bound state in the
$K^- pp$ system on the ''less important'' interactions, which are the interactions in the lowest channels.
We also wrote and solved the three-body Faddeev-type AGS equations with three coupled $\bar{K}NN$, $\pi \Sigma N$,
and $\pi \Lambda N$ channels and directly included our new antikaon-nucleon potentials with coupled
$\bar{K}N - \pi \Sigma - \pi \Lambda$ channels. It was shown in \cite{FineTune_FBS} that $YN$
interaction model can have a visible effect on the three-body result, while $\pi N$ potential can change the result
only slightly. The three-body AGS equations with three coupled channels strongly changed the $K- pp$ characteristics
in comparison with those obtained from solving AGS equations with two coupled channels.

In the present paper, we continued the study started in \cite{FineTune_FBS}.  Namely, we
constructed accurate  $\pi N$ potentials and together with our best $YN$ model from \cite{FineTune_FBS}  used them
in the three-body calculations of the $K^- pp$  quasi-bound state. The new antikaon-nucleon potentials with three coupled
particle channels were refitted and directly included in the three-body AGS equations
with three coupled $\bar{K}NN$, $\pi \Sigma N$, and $\pi \Lambda N$ channels. 

We did not focus on the $K^- pp$ system only. Another state of the $\bar{K}NN$ system, namely $K^- pn$
with a possible quasi-bound state caused by the strong interactions \cite{Kd_qbs} was also studied using
the same new antikaon-nucleon, hyperon-nucleon, pion-nucleon potentials, and the three-body formalism with
three coupled channels. Kaonic deuterium, which is another possible quasi-bound state in the $K^- np$ system
caused mainly by Coulomb interaction, is not considered in this paper.
Finally, the binding energy and width of the three-body system with two antikaons $\bar{K} \bar{K}N$ studied by
us in ~\cite{KKN} were recalculated using the new $\bar{K}N - \pi \Sigma - \pi \Lambda$ potentials. For this system,
we also derived and solved three-body Faddeev-type AGS equations with three coupled
$\bar{K} \bar{K} N$, $\bar{K} \pi \Sigma$, and  $\bar{K} \pi \Lambda$ channels. 

The paper is organized as follows. The three-body Faddeev-type AGS equations with three coupled channels 
are described in the next section. Section~ \ref{TwoBodyInput_sect} contains information on the two-body
input for the few-body equations. First, the new antikaon-nucleon interaction models with three coupled $\bar{K}N$,
$\pi \Sigma$, and $\pi \Lambda$ channels are described. The next subsection is devoted to the nucleon-nucleon
and antikaon-antikaon potentials. In the following subsections, the new hyperon-nucleon and
pion-nucleon interaction models are described. 
Section~\ref{results_sect} is devoted to the results of the three-body calculations.
Binding energies and widths of the $K^- pp$ and $K^- np$ systems, obtained from the three-body
coupled-channel calculations with new  $\bar{K}N - \pi \Sigma - \pi \Lambda$, $YN$, and $\pi N$ potentials 
are presented and discussed in Subsections~\ref{Kpp_res.sect} and \ref{Kpn_res.sect}, correspondingly. 
Subsection~\ref{KKN_res.sect} contains predictions of the $\bar{K} \bar{K} N$ quasi-bound state
characteristics, while the last subsection of \ref{results_sect} contains the joint three-body results.  The paper is
finished by the Conclusions section.

\section{Three-body Faddeev-type AGS equations with three coupled particle channels}
\label{Few-bodyEqs_sect}

Dynamically exact Faddeev-type AGS equations  \cite{AGS3} with coupled particle channels were used
by us for calculating characteristics of the quasi-bound states in the $\bar{K}NN$ and $\bar{K}\bar{K}N$ systems.
It turned out that these equations written in momentum representation are the most suitable ones for the study of systems consisting of antikaon(s) and nucleon(s). First, Faddeev-type equations in momentum
representation allow us to calculate the width of the quasi-bound state directly as the doubled imaginary
part of the three-body pole corresponding to the quasi-bound state. It is impossible to calculate it accurately in
coordinate representation. Next, the coupled particle channels, which, as was demonstrated in our previous
three-body calculations \cite{my_review}, are very important for the few-body systems with antikaon(s) and
nucleon(s), can also be introduced directly into the Faddeev-type equations. The different energies
in the channels can be naturally defined in the kernels of the integral equations. Finally, only Faddeev-type equations in momentum representation can treat energy-dependent interaction models exactly. It is the case
of one of our potentials, the chirally motivated one, which is assumed to be the most advanced models nowadays.

We used one- and two-term separable potentials for all two-body interactions as an input. If separable potentials
${V}_{i}$ are used, the corresponding $T$-matrices ${T}_{i}$ are also separable:
\begin{equation}
\label{VTsep}
{V}_{i} =  \lambda_i \, | {g}_{i} \rangle  \langle {g}_{i} |, \;
{T}_{i} =  | {g}_{i} \rangle  \,  {\tau}_{i} \,   \langle {g}_{i} |.
\end{equation}
Here a one-term separable potential is shown for simplicity. The Faddeev-type AGS equations for such potentials
in the operator form
\begin{equation}
\label{AGS3ch}
 {X}_{ij}^{\alpha \beta}(z) = \delta_{\alpha \beta} {Z}_{ij}^{\alpha} + 
 \sum_{k=1}^3 \sum_{\gamma=1}^5  {Z}_{ik}^{\alpha} \, {\tau}_{k}^{\alpha \gamma} \,
   {X}_{kj}^{\gamma \beta}
\end{equation}
contain three-body transition ${X}_{ij}^{\alpha \beta}$ and kernel ${Z}_{ij}^{\alpha}$ operators.
Faddeev indices $i,j,k = 1,2,3$ in Eq.(\ref{AGS3ch}) simultaneously denote a pair of particles and
the third particle, a spectator. The additional indices $\alpha, \beta, \gamma$
denote a particle channel. It is known that antikaon-nucleon interaction is coupled
to $\pi \Sigma$ and $\pi \Lambda$ channels. Only the $\pi \Sigma$ channel,
which is the channel where $\Lambda(1405)$ resonance is formed,
was directly included into the three-body equations in our previous calculations.
Now we directly included the lowest $\pi \Lambda$ channel as well. It means that
the AGS equations Eq.(\ref{AGS3ch}) for the $\bar{K}NN$ system contain the following five three-body channels
\begin{widetext}
\begin{eqnarray}
\label{KNNchnls}
\alpha = 1: | \bar{K}_1N_2N_3 \rangle, \qquad  \alpha = 2: | \pi_1 \Sigma_2 N_3 \rangle, 
\qquad \alpha = 3: | \pi_1 N_2 \Sigma_3 \rangle, \\
\nonumber
{ \alpha = 4: | \pi_1 \Lambda_2 N_3 \rangle,  \qquad
\alpha = 5: | \pi_1 N_2 \Lambda_3 \rangle},
\end{eqnarray}
\end{widetext}
while for the $\bar{K}\bar{K} N$ system the channels are
\begin{widetext}
\begin{eqnarray}
\label{KKNchnls}
\alpha = 1: | \bar{K}_1\bar{K}_2N_3 \rangle,  \qquad \alpha = 2: | \bar{K}_1 \pi_2 \Sigma_3 \rangle,
\qquad \alpha = 3: | \pi_1 \bar{K}_2 \Sigma_3 \rangle, \\
\nonumber
{ \alpha = 4: | \bar{K}_1 \pi_2 \Lambda_3 \rangle, \qquad
 \alpha = 5: | \pi_1 \bar{K}_2 \Lambda_3 \rangle}.
\end{eqnarray}
\end{widetext}

The system of operator equations Eqs.(\ref{AGS3ch}) for the $\bar{K}NN$ system was written down in momentum
representation and antisymmetrized due to identical fermions in the highest $\bar{K}NN$ channel.
The AGS equations for the $\bar{K}\bar{K}N$ system were symmetrized due to identical baryons in the highest channel.
The resulting systems of the coupled integral equations were solved numerically. 

The three-body $\bar{K}NN$ systems with spin $S^{(3)}_{\bar{K}NN} = 0$ (corresponding to the $K^- pp$ in particle
representation) and spin $S^{(3)}_{\bar{K}NN}  = 1$ (corresponding to the $K^- np$) and  isospin $I^{(3)}_{\bar{K}NN} =1/2$
were studied. The total spin of the $\bar{K} \bar{K} N$ system is one-half, and the isospin was also chosen to be
$I^{(3)}_{\bar{K}\bar{K}N} =1/2$.

\section{Two-body input}
\label{TwoBodyInput_sect}

The input for the three-body Faddeev-type equations Eq.(\ref{AGS3ch}) are two-body
$T$-matrices corresponding to the potentials Eq.(\ref{VTsep}), describing two-body interactions between
every pair of particles in all particle channels. For the $\bar{K}NN$ and $\bar{K} \bar{K} N$ three-body 
systems $V_{\bar{K}N - \pi \Sigma - \pi \Lambda}$, $V_{NN}$, $V_{\bar{K} \bar{K}}$, $V_{\Sigma N - \Lambda N}$,
and $V_{\pi N}$ interaction models should be known. As before, we assumed that $\bar{K} \pi$ and $\bar{K} Y$
interactions in the $\bar{K} \bar{K} N$ system are not important and neglected them.
All our potentials are separable $s$-wave ones.

\subsection{Antikaon-nucleon potentials with $\bar{K}N - \pi \Sigma -\pi \Lambda$ coupled channels}
\label{VKN}

The antikaon-nucleon interaction plays the most important role in the three-body
calculations of the considered systems. The corresponding results strongly depend on it, as was
demonstrated by us in \cite{Kd_Kpp_last} and earlier papers.
Antikaon-nucleon potentials with three coupled $\bar{K}N$, $\pi \Sigma$, and $\pi \Lambda$
channels are necessary for solving the system of integral AGS equations Eqs.(\ref{AGS3ch}) with
three coupled three-body channels.

Three models of the antikaon-nucleon interaction were constructed and used by us in the past 
in our studies of three- and four-body systems consisting of antikaon(s) and nucleon(s).
Two of them are phenomenological potentials with one- or two-pole structure
of the $\Lambda(1405)$ resonance~\cite{VKN_SIDD12}, which plays
an important role in the interaction, coupling $\bar{K}N$ with $\pi \Sigma$ channel. 
\begin{table*}
\begin{center}
\begin{tabular}{ccccccc}
\hline  \noalign{\smallskip}
  & $V_{\bar{K}N-\pi \Sigma - \pi \Lambda}^{\rm 1,SIDD-A}$  &  $V_{\bar{K}N-\pi \Sigma - \pi \Lambda}^{\rm 1,SIDD-B}$  &
  $V_{\bar{K}N-\pi \Sigma - \pi \Lambda}^{\rm 2,SIDD-A}$  &  $V_{\bar{K}N-\pi \Sigma - \pi \Lambda}^{\rm 2,SIDD-B}$  &
    $V_{\bar{K}N-\pi \Sigma - \pi \Lambda}^{\rm Chiral}$  & Exp\\[1mm]
\noalign{\smallskip} \hline \noalign{\smallskip}
 $\Delta E_{1s}$   & $-314.1$ & $-306.4$ & $-293.8$ & $-292.5$ & $-319.9$ & $-283\pm 36 \pm 6$ \\[1mm]
 $\Gamma_{1s}$  & $626.9$ & $600.7$ &  $639.8$ & $609.8$ & $622.0$ & $541 \pm 89 \pm 22$ \\[1mm]
 $E_1$ & $1429.2 - i \, 33.5$ & $1427.6 - i \, 36.9$  & $1428.2 - i \, 40.9$ & $1426.8 - i \, 45.8$ & $1424.5 - i \, 29.0$ & $-$\\[1mm]
 $E_2$ & $-$ & $-$ &  $1355.3 - i \, 95.6$ &  $1353.1 - i \, 87.8$ & $1376.1 - i \, 84.1$ & $-$\\[1mm]
 $\gamma$ & $2.36$ &  $2.35$ & $2.36$ &  $2.36$ & $2.36$ & $2.36 \pm 0.04$ \\[1mm]
 $R_c$         & $0.655$ & $0.657$ & $0.654$ & $0.654$ & $0.656$ & $0.664 \pm 0.011$ \\[1mm]
 $R_n$        & $0.188$ & $0.191$ & $0.185$ & $0.189$ & $0.187$ & $0.189 \pm 0.015$ \\[1mm]
 $a_{K^- p}$   & $-0.75 +i \, 0.94$ & $-0.74 +i \, 0.89$ & $-0.69 +i \, 0.94$ & $-0.70 +i \, 0.90$ & $-0.77 +i \, 0.94$ &  $-$\\[1mm]
 \noalign{\smallskip} \hline
\end{tabular}
\caption{Physical characteristics given by the new phenomenological $\bar{K}N -  \pi \Sigma - \pi \Lambda$ potentials:
$1s$ level shift $\Delta E_{1s}$ (eV) and width  $\Gamma_{1s}$ (eV), strong poles $E_1$ (MeV) and $E_2$ (MeV), 
threshold branching ratios $\gamma$, $R_c$, $R_n$. Calculations were performed with physical masses in 
$\bar{K}N$ channel and Coulomb interaction in the $K^- p$ pair. The $K^- p$ scattering length $a_{K^- p}$ (fm) is
also shown. A and B versions of the phenomenological one- $V_{\bar{K}N-\pi \Sigma - \pi \Lambda}^{\rm 1,SIDD}$ and 
two-pole $V_{\bar{K}N-\pi \Sigma - \pi \Lambda}^{\rm 2,SIDD}$ potentials are those with negative or posiive
$I=1$ strength constants, correspondingly. Experimental data on
characteristics of kaonic hydrogen \cite{SIDDHARTA} and threshold branching ratios \cite{gammaKp1,gammaKp2}
are presented for comparison. 
\label{KNobserv.tab}
}
\end{center}
\end{table*}

A one-term separable potential Eq.~(\ref{VTsep}) with
coupled particle channels written in momentum representation has a form
\begin{equation}
  V_{I}^{\alpha \beta}(k^{\alpha},{k'}^{\beta}) = 
   \lambda_{I}^{\alpha \beta} \, g(k^{\alpha}) g({k'}^{\beta}),
\label{VSep}
\end{equation}
where $\alpha, \beta$ are indices of the two-body channels, $I$ is a two-body isospin.
The resonance appears there as a quasi-bound state in the $\bar{K}N$ ($\alpha, \beta = 1$) and as
a resonance in the lower $\pi \Sigma$ ($\alpha,\beta = 2$)
channel. The form factors of the one-pole version of the
phenomenological potential $V_{\bar{K}N}^{\rm 1,SIDD}$ and those of the $\bar{K}N$ channel of
the two-pole version $V_{\bar{K}N}^{\rm 2,SIDD}$ have a Yamaguchi form
\begin{equation}
 g_I^{\alpha} = \frac{1}{(k^{\alpha})^2 + (\beta_I^{\alpha})^2},
\label{YamagFF}
\end{equation}
while for the $\pi \Sigma$ channel  in the two-pole model of the interaction it has the following form
\begin{equation}
 g_I^{\alpha} = \frac{1}{(k^{\alpha})^2 + (\beta_I^{\alpha})^2} +
  \frac{s \, (\beta_I^{\alpha})^2}
        {\left[ (k^{\alpha})^2 + (\beta_I^{\alpha})^2 \right]^2}.
\end{equation}
The $\pi \Lambda$ channel had been taken in the phenomenological potentials 
$V_{\bar{K}N}^{\rm 1,SIDD}$ and $V_{\bar{K}N}^{\rm 2,SIDD}$ \cite{VKN_SIDD12} into account indirectly through
imaginary part of the complex strength parameter $\lambda^{11}_{I=1}$.

One more previously constructed and used antiakon-nucleon interaction model is a chirally motivated potential 
$V_{\bar{K}N}^{\rm Chiral}$ \cite{ourKNN_I} which couples $\bar{K}N$,  $\pi \Sigma$, and $\pi \Lambda$ channels. 
In contrast to the energy-independent phenomenological models, the chirally motivated potential has energy-dependent
strength parameters $\lambda^{\alpha \beta}_{I}$.

Recently \cite{FineTune_FBS}, we constructed new versions of the phenomenological potentials that directly couple
all three particle channels $\bar{K}N$, $\pi \Sigma$, and $\pi \Lambda$.
As before, the $V^{\rm 1,SIDD}_{\bar{K}N - \pi \Sigma - \pi \Lambda}$ interaction model
has one pole corresponding to the $\Lambda(1405)$ resonance, while 
the $V^{\rm 2,SIDD}_{\bar{K}N - \pi \Sigma - \pi \Lambda}$ potential lead to two
strong poles. Yamaguchi form-factors Eq.(\ref{YamagFF}) are used for the $\pi \Lambda$ channel ($\alpha, \beta = 3$
in Eq.(\ref{VSep})) in the both versions. We also refitted parameters of the chirally motivated potential
to set the strong poles corresponding to the $\Lambda(1405)$ resonance closer to other chiral models.
\begin{table*}[ht]
\begin{center}
\begin{tabular}{ccccc}
\hline  \noalign{\smallskip}
  & $V_{\bar{K}N-\pi \Sigma - \pi \Lambda}^{\rm 1,SIDD-A}$   & $V_{\bar{K}N-\pi \Sigma - \pi \Lambda}^{\rm 1,SIDD-B}$  
  & $V_{\bar{K}N-\pi \Sigma - \pi \Lambda}^{\rm 2,SIDD-A}$   & $V_{\bar{K}N-\pi \Sigma - \pi \Lambda}^{\rm 2,SIDD-B}$
     \\[1mm]
\noalign{\smallskip} \hline
$\beta_1$ & $3.95$ & $3.86$ &  $4.00$ & $3.76$ \\
$\beta_2$ & $2.06$ & $1.96$ &  $1.38$ & $1.20$ \\
$\beta_3$ & $1.42$ & $0.50$ & $2.21$ & $0.50$  \\
$s$ & 0. & 0. & $-0.70$ & $-0.70$ \\
$\lambda_{11,0}$ &   $-1.96$ & $-1.85$  & $-1.95$ & $-1.64$  \\
$\lambda_{12,0}$ &   $0.70$ & $0.67$  & $0.68$ & $0.56$ \\
$\lambda_{22,0}$ &   $-0.06$ & $-0.01$  & $-0.19$ & $-0.05$  \\
$\lambda_{11,1}$ &   $-0.21$ & $0.66$  & $-0.11$ & $0.90$  \\
$\lambda_{12,1}$ &  $1.11$ & $2.00$  & $1.19$ & $2.00$  \\
$\lambda_{22,1}$ &   $-2.2E-8$ & $1.48$ & $-5.5E-5$ & $1.46$   \\
$\lambda_{13,1}$ &   $0.69$ & $0.41$  & $1.30$ & $0.47$  \\
$\lambda_{23,1}$ &   $1.13$ & $0.47$  & $1.95$ & $0.50$ \\
$\lambda_{33,1}$ &   $-0.11$ & $0.13$  & $-0.59$ & $0.15$ \\[1mm]
 \noalign{\smallskip} \hline
\end{tabular}
\caption{Parameters of the phenomenological antikaon-nucleon potentials with coupled
$\bar{K} N - \pi \Sigma - \pi \Lambda$ channels. The range $\beta_{\alpha}$ (fm$^{-1}$), strength
$\lambda_{\alpha \beta, I}$ parameters, and an additional parameter $s$ for the two-pole
version are presented ($\alpha, \beta = 1,2,3$ denote the $\bar{K}N$, $\pi \Sigma$, and $\pi \Lambda$
channels respectively; $I$ is a two-body isospin).  A and B versions of the phenomenological
one- $V_{\bar{K}N-\pi \Sigma - \pi \Lambda}^{\rm 1,SIDD}$ and 
two-pole $V_{\bar{K}N-\pi \Sigma - \pi \Lambda}^{\rm 2,SIDD}$ potentials are those with negative
or positive $I=1$ strength constants, correspondingly.
\label{KNparams_phen.tab}
}
\end{center}
\end{table*}
%
\begin{table*}[ht]
\begin{center}
\begin{tabular}{ccccccc}
\hline  \noalign{\smallskip}
  $f_{\pi}$ & $f_{K}$   & $\beta_{1,0}$ & $\beta_{2,0}$ & $\beta_{1,1}$ & $\beta_{2,1}$ & $\beta_{3,1}$  
     \\[1mm]
\noalign{\smallskip} \hline
\; $111.90$ \; & \;  $108.01$ \;  & \; $3.99$ \;  & \; $2.94$ \;  & \; $2.96$ \;  & \;  $4.04$ \;  & \;  $4.85$ \;  \\[1mm]
 \noalign{\smallskip} \hline
\end{tabular}
\caption{Parameters of the chirally motivated antikaon-nucleon potential with coupled
$\bar{K} N - \pi \Sigma - \pi \Lambda$ channels. The pseudo-scalar meson decay constants 
$f_{\pi}$, $f_{K}$ (MeV) and range parameters $\beta_{\alpha, I}$ (fm$^{-1}$) are presented
($\alpha = 1,2,3$ denote the $\bar{K}N$, $\pi \Sigma$, and $\pi \Lambda$
channels respectively; $I$ is a two-body isospin).
\label{KNparams_chiral.tab}
}
\end{center}
\end{table*}

In the present study we performed new fits of our antikaon-nucleon potentials with additional conditions.
Namely, we put constraints on some of the strength constants of the phenomenological models. 
In addition to the condition of the negative value of the $\lambda^{11}_{0}$ constant describing isospin
zero $\bar{K}N,\bar{K}N$ element of the total strength matrix (with elements $\lambda^{\alpha \beta}_I$)
we also forced $\lambda^{22}_{0}$ for the $\pi \Sigma,\pi \Sigma$ element to be negative as well. It means
that the corresponding parts of the interaction are attractive. 

Recently, the E15 collaboration reported results of their measurements of the mesonic decay branch of the
$\bar{K}NN$ quasibound state \cite{E15_JPARC3} . A $I_{\bar{K}N} = 1$ cusp, which was observed in differential
cross-sections. led to a suggestion, that ''the real part of the $I_{\bar{K}N} = 1$ interaction is also attractive,
although it is not sufficiently strong to form a bound state in this channel''. In order to check this hypothesis,
we put an additional constraint on the diagonal $\lambda^{\alpha \alpha}_{I=1}$ constants to be negative (attractive).
The best fits to two-body observables were obtained when all three diagonal strength constants, for
the $\bar{K}N$, $\pi \Sigma$, and $\pi \Lambda$ channels, are negative. 
The phenomenological potentials with these sets of parameters are denoted as A versions:
$V_{\bar{K}N-\pi \Sigma - \pi \Lambda}^{\rm 1,SIDD-A}$ and $V_{\bar{K}N - \pi \Sigma - \pi \Lambda}^{\rm 2,SIDD-A}$.
Potentials with another sets of parameters for the phenomenological potentials without this additional condition on
$\lambda^{\alpha \beta}_{I=1}$ constants (which tend to be positive) are denoted as B versions
$V_{\bar{K}N-\pi \Sigma - \pi \Lambda}^{\rm 1,SIDD-A}$ and $V_{\bar{K}N - \pi \Sigma - \pi \Lambda}^{\rm 2,SIDD-A}$.
 Since $\lambda^{\alpha \beta}_{I}$ parameters of the chirally
motivated potential are energy dependent functions, no such conditions can be put on them. Due to this, only one
version of the chirally motivated potential was refitted.
%
\begin{figure*}
 \begin{center}
 \includegraphics[width=0.65\textwidth]{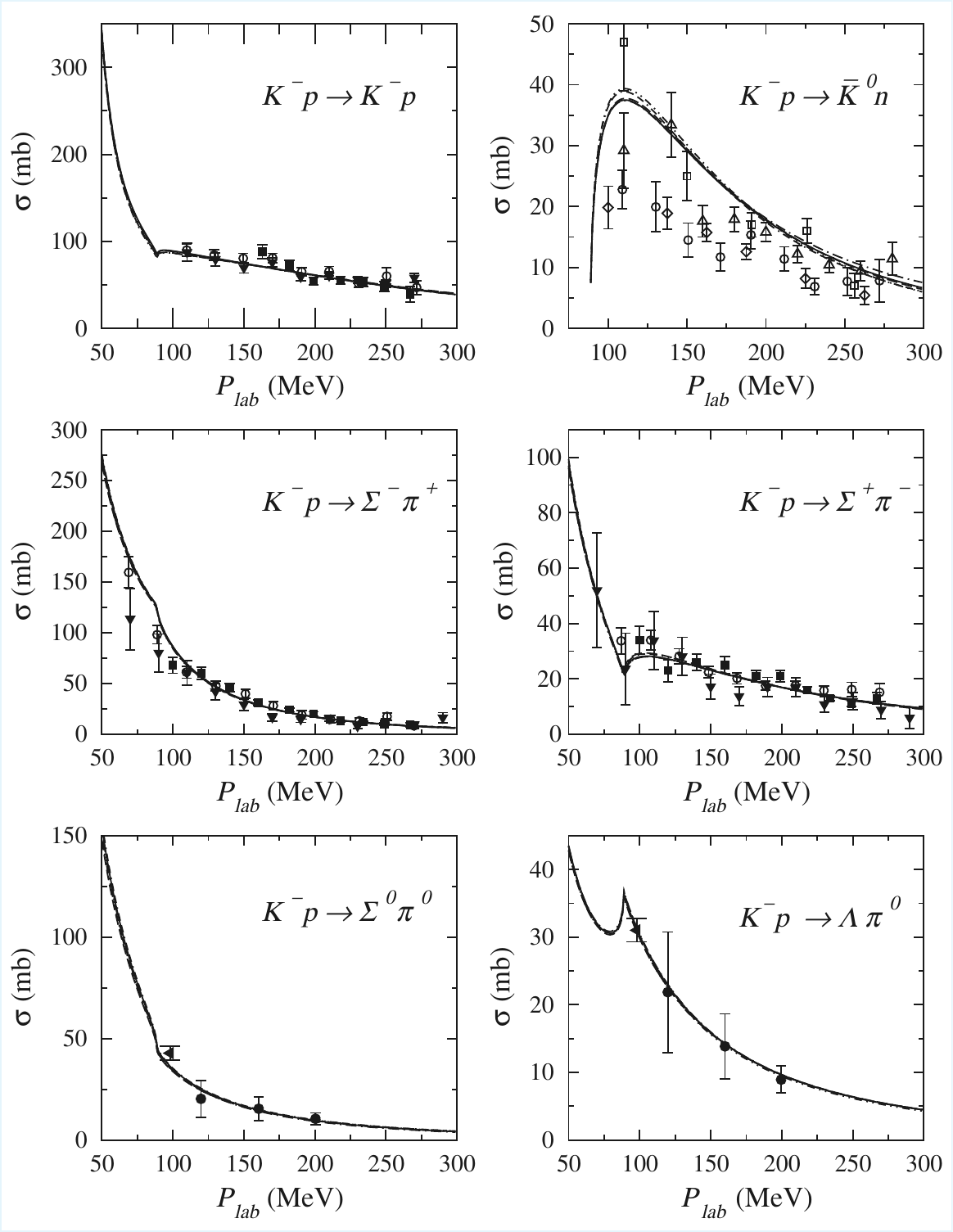}
 \end{center}

 \caption{$K^- p$ elastic and inelastic low-energy cross-sections given by our antikaon-nucleon
potentials coupling $\bar{K}N$, $\pi \Sigma$, and $\pi \Lambda$ channels.  Straingt line (denoting results
calculated with  $V^{\rm 1,SIDD-A}_{\bar{K}N - \pi \Sigma - \pi \Lambda}$ potential),
long dashed line ($V^{\rm 1,SIDD-B}_{\bar{K}N - \pi \Sigma - \pi \Lambda}$),
dash-dash-dot line ($V^{\rm 2,SIDD-A}_{\bar{K}N - \pi \Sigma - \pi \Lambda}$), 
dot-dot-dash line ($V^{\rm 2,SIDD-B}_{\bar{K}N - \pi \Sigma - \pi \Lambda}$),
and dash-dot line
($V^{\rm Chiral}_{\bar{K}N - \pi \Sigma - \pi \Lambda}$) are compared with experimental data \cite{Kp2exp,Kp3exp_1,Kp3exp_2,Kp4exp,Kp5exp,Kp6exp,KpLASTexp} (symbols).
A and B versions of the phenomenological one-  and two-pole potentials are those with negative or positive
$I=1$ strength constants, correspondingly.
}
\label{Kp_CrosSect.fig}
\end{figure*}

The parameters of the new phenomenological and the chirally motivated potentials were, as before, fitted
to the experimental data on elastic  and inelastic $K^- p$ cross-sections. However, now we chose only
two experimental data sets for each of the $K^- p \to K^- p$ and inelastic $K^- p \to \pi^+ \Sigma^-$, 
$K^- p \to \pi^- \Sigma^+$, and $K^- p \to \pi^0 \Sigma^0$, $K^- p \to \pi^0 \Lambda$ 
reactions out of the existing ones~\cite{Kp2exp,Kp3exp_1,Kp3exp_2,Kp4exp,Kp5exp,Kp6exp}.
Since the data on the inelastic $K^- p \to \bar{K}^0 n$ cross-sections from different experiments
strongly contradict each other, we did not use them now in our fits. Nevertheless, the cross-sections given by all 
our new potentials reproduce some of the experimental data sets on $K^- p \to \bar{K}^0 n$ scattering very well
without being fitted to them. One more difference from our previous fits is that we additionally fitted
the potential parameters  to the recently measured accurate data on $K^- p \to \pi^0 \Sigma^0$ and
$K^- p \to \pi^0 \Lambda$ cross-sections \cite{KpLASTexp}.

All new antikaon-nucleon potentials now reproduce the threshold branching ratios
$\gamma$, $R_c$ and $R_n$~\cite{gammaKp1, gammaKp2} directly
\begin{widetext}
\begin{eqnarray}
\label{gamma}
\gamma &=& \frac{\Gamma(K^- p \to \pi^+ \Sigma^-)}{\Gamma(K^- p \to
\pi^- \Sigma^+)} = 2.36 \pm 0.04, \\
\label{Rc}
R_c &=& \frac{\Gamma(K^- p \to \pi^+ \Sigma^-, \pi^- \Sigma^+)}{\Gamma(K^- p \to
\mbox{all inelastic channels} )}   = 0.664 \pm 0.011, \\
\label{Rn}
R_n &=& \frac{\Gamma(K^- p \to \pi^0 \Lambda)}{\Gamma(K^- p \to
\mbox{neutral states} )} = 0.189 \pm 0.015.
\end{eqnarray}
\end{widetext}
All $\bar{K}N$ interaction models, similar to our previous versions, reproduce the most recent experimental results
by SIDDHARTA collaboration~\cite{SIDDHARTA} on the $1s$ level shift $\Delta E_{1s}$ and width $\Gamma_{1s}$
of kaonic hydrogen
\begin{eqnarray}
\Delta E_{1s}  &=&  -283 \pm 36 \pm 6 \; {\rm eV}, \\
\Gamma_{1s}  &=&  541 \pm 89 \pm 22 \; {\rm eV}.
\end{eqnarray}
The energy of the $1s$ level and the width were calculated by us directly by solving the Lippmann-Schwinger
equation with one of the strong $\bar{K}N - \pi \Sigma - \pi \Lambda$ hadronic interaction model plus Coulomb
potential for the $K^- p$ pair.  The same equations with strong interactions together with Coulomb interactions were solved
during fitting cross-sections and threshold branching ratios. It is the unique property of our
antikaon-nucleon potentials, including the chirally motivated one.  As far as we know, none of the other $\bar{K}N$
interaction models was constructed with the Coulomb interaction directly included in the equations.
In addition, the physical masses of the particles in the highest $\bar{K}N$ channel were used in the calculations
of the observables.  On the contrary, the three-body calculations were performed with isospin-averaged masses
and without Coulomb interaction, since we assume these effects play in this case a minor role.

All our new $\bar{K}N-\pi \Sigma - \pi \Lambda$ potentials 
(A- or B-versions of the phenomenological ones with one- or two-pole structure of the $\Lambda(1405)$ resonance
and the chirally motivated one) describe the experimental data with the same level of accuracy. It can be seen in Table \ref{KNobserv.tab} and in Fig. \ref{Kp_CrosSect.fig}, where the cross-sections given by the new antikaon-nucleon
potentials with three coupled channels are shown together with the experimental data. The data used in the fits
of the potential parameters are denoted in Fig. \ref{Kp_CrosSect.fig} as filled symbols, the remaining ones as empty symbols.
The only small differences between the theoretical results are seen in the $K^- p \to \bar{K}^0 n$ cross-sections.
In addition, the isospin-zero
elastic $\pi \Sigma$ cross sections have a single peak near the mass of the $\Lambda(1405)$ resonance
($M_{\Lambda(1405)} = 1405.1$ MeV according to PDG~\cite{PDG}) irrespective of the number of corresponding
poles produced by the particular version of the antikaon-nucleon potential. The parameters of the phenomenological
potentials are presented in Table \ref{KNparams_phen.tab}, those of the chiral potential - in
Table  \ref{KNparams_chiral.tab}.

\subsection{Nucleon-nucleon and antikaon-antikaon interactions}
\label{VNN_KK}

Nucleon-nucleon potential is necessary for calculations of the $\bar{K}NN$ system.
Dependence of the three-body results on the nucleon-nucleon interaction models
was studied in our previous works, see e.g.~\cite{my_review}. It turned out that $V_{NN}$ is important 
only for the $K^- np$ system, where the particular version of the potential can resolve the question
of quasi-bound state existence. The reason is that the quasi-bound state in $K^- np$
caused by strong interactions only (there is also an atomic state there, a kaonic deuterium)
is situated very close to the threshold. For other few-body systems consisting of antikaon and
nucleons, the nucleon-nucleon interaction plays a minor role.

The best of the checked nucleon-nucleon models,  the Two-term Separable New potential (TSN)
constructed in \cite{Kd_Kpp_last} was used in the present study. It reproduces phase shifts of Argonne $V18$
potential \cite{ArgonneV18},  triplet and singlet scattering lengths and deuteron
binding energy. Parameters and physical characteristics given by the potential can be
found in \cite{Kd_Kpp_last}.

Another interaction, namely antikaon-antikaon one, is necessary for the investigation of the $\bar{K} \bar{K} N$
system. Experimental information on the interaction is absent, so a model based
on a $\pi \pi - \bar{K}K$ J\"ulich interaction was constructed first.
Phase shifts given by the $\bar{K}\bar{K}$ model were then used for fitting parameters of two versions
of the phenomenological $\bar{K} \bar{K}$ potentials \cite{KKN}. In the present study, we used the
''Lattice motivated'' version of the phenomenological $\bar{K}\bar{K}$ potential. A more detailed
description of the antikaon-antikaon potential can be found in \cite{KKN}.

\subsection{$\Sigma N - \Lambda N$ interaction models}
\label{VYN}

Knowledge of hyperon-nucleon interaction is necessary for the study of systems consisting
of an antikaon and two nucleons.
The $YN$ interaction strongly depends on the two-body isospin of the system: 
isospin $I=1/2$  $\Sigma N$ system is coupled to the $\Lambda N$ channel, 
while $I=3/2$  $\Sigma N$ system stays alone.
A simple separable model of the $\Sigma N - \Lambda N$ potential was constructed and
used in our previous calculations \cite{Kd_Kpp_last} with parameters  fitted to experimental
cross-sections only. Recently \cite{FineTune_FBS}, we checked the dependence of the three-body results
on  the $YN$ interaction model. 
Four versions of the potential: spin-dependent or spin-independent ones, fitted
either to experimental cross-sections~\cite{SigmaN1,SigmaN2,SigmaN3,SigmaN4,SigmaN5}
only or to the cross-sections together with scattering lengths of different states of
$\Sigma N$ and $\Lambda N$ systems given by an "advanced" potential \cite{HyperN_poten} were used.
It turned out  that the dependence of the three-body $K^- pp$ results on the $YN$ model can be quite strong.
The present calculations were performed with the best of the four versions of the
potential \cite{FineTune_FBS}, which is the spin-dependent one with parameters fitted to both experimental
cross-sections and "advanced" scattering lengths.

\begin{figure*}
 \begin{center}
 \includegraphics[width=0.40\textwidth]{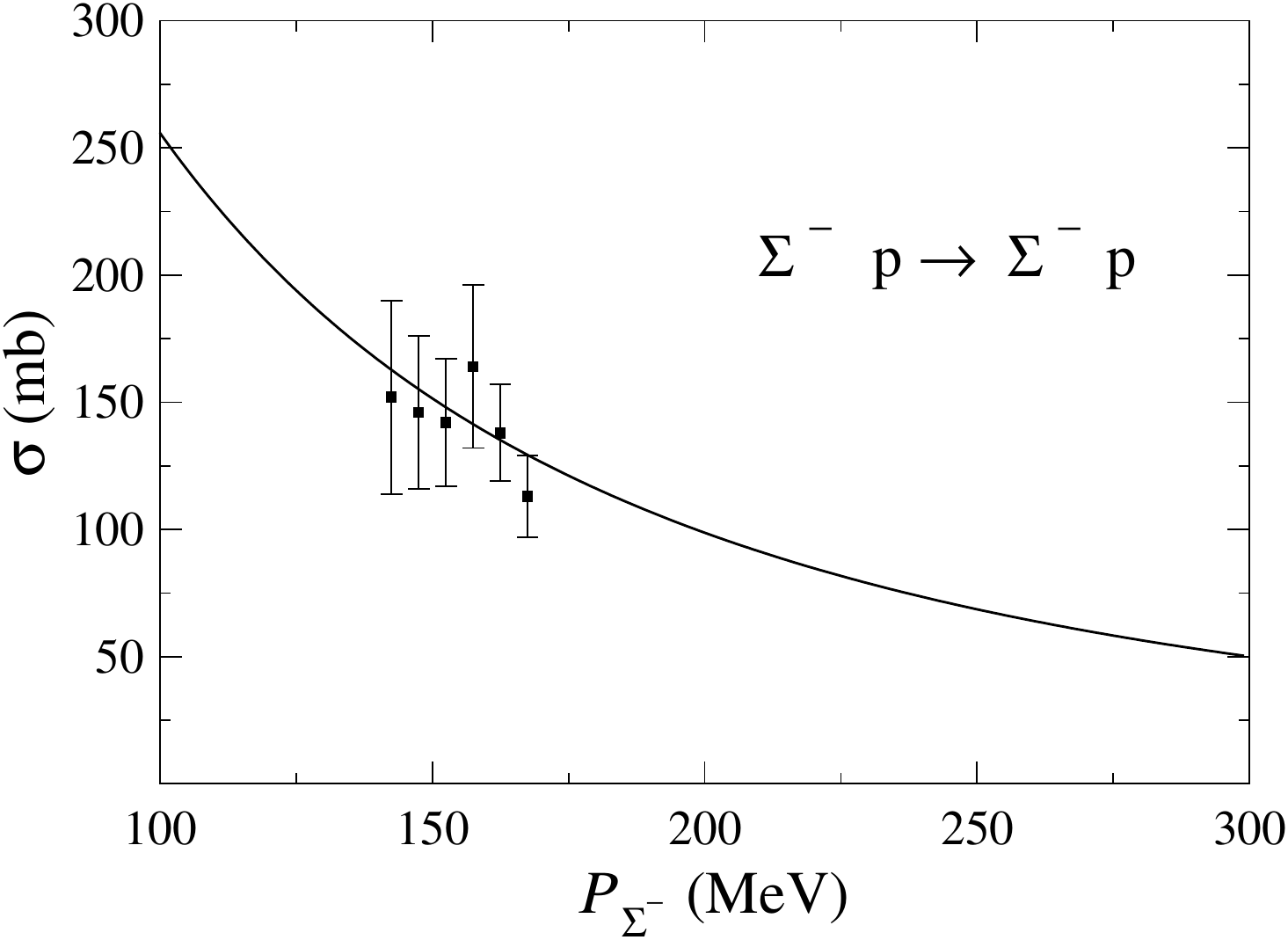}
\includegraphics[width=0.40\textwidth]{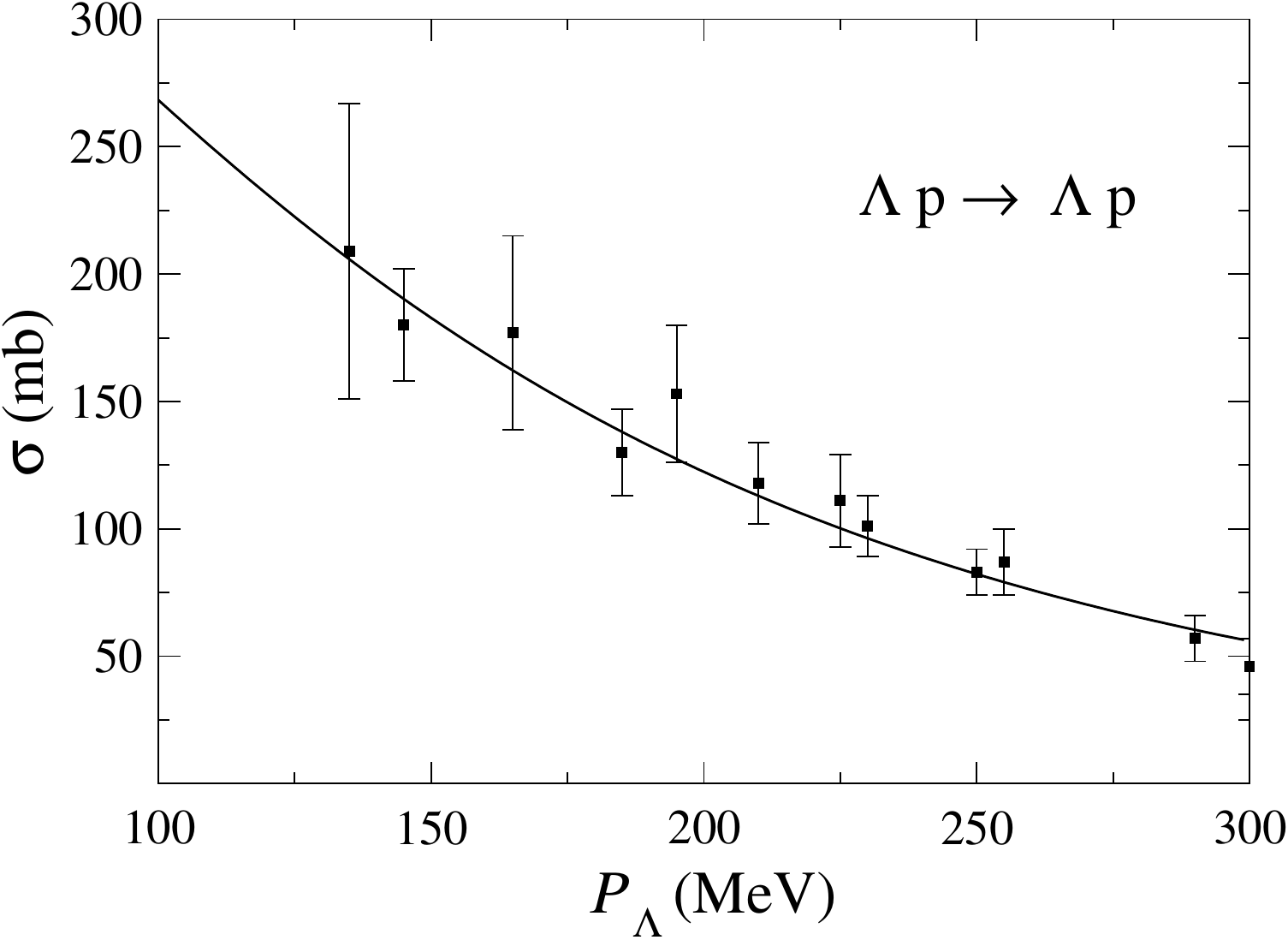} \\[3mm]
\includegraphics[width=0.40\textwidth]{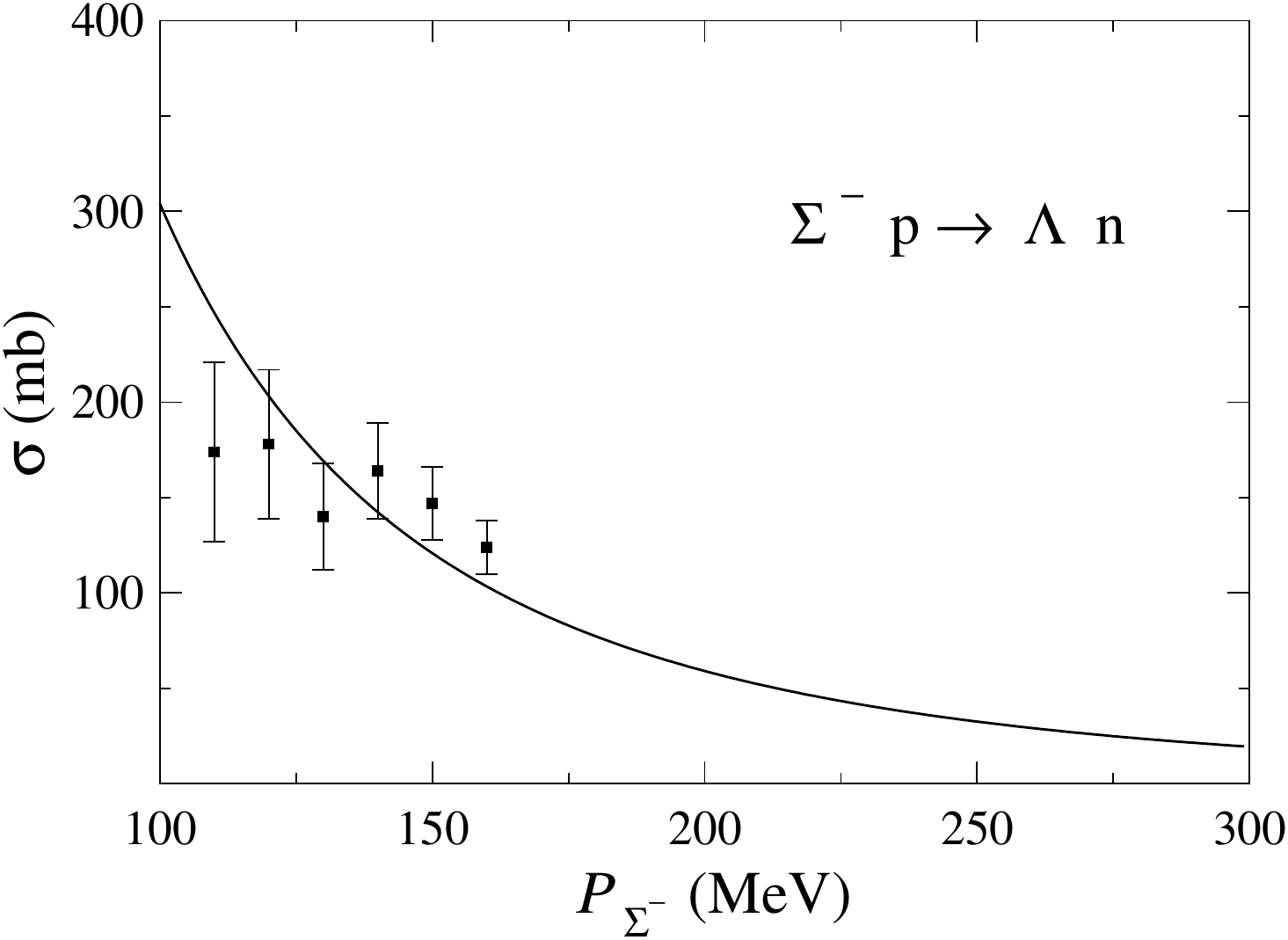}
\includegraphics[width=0.40\textwidth]{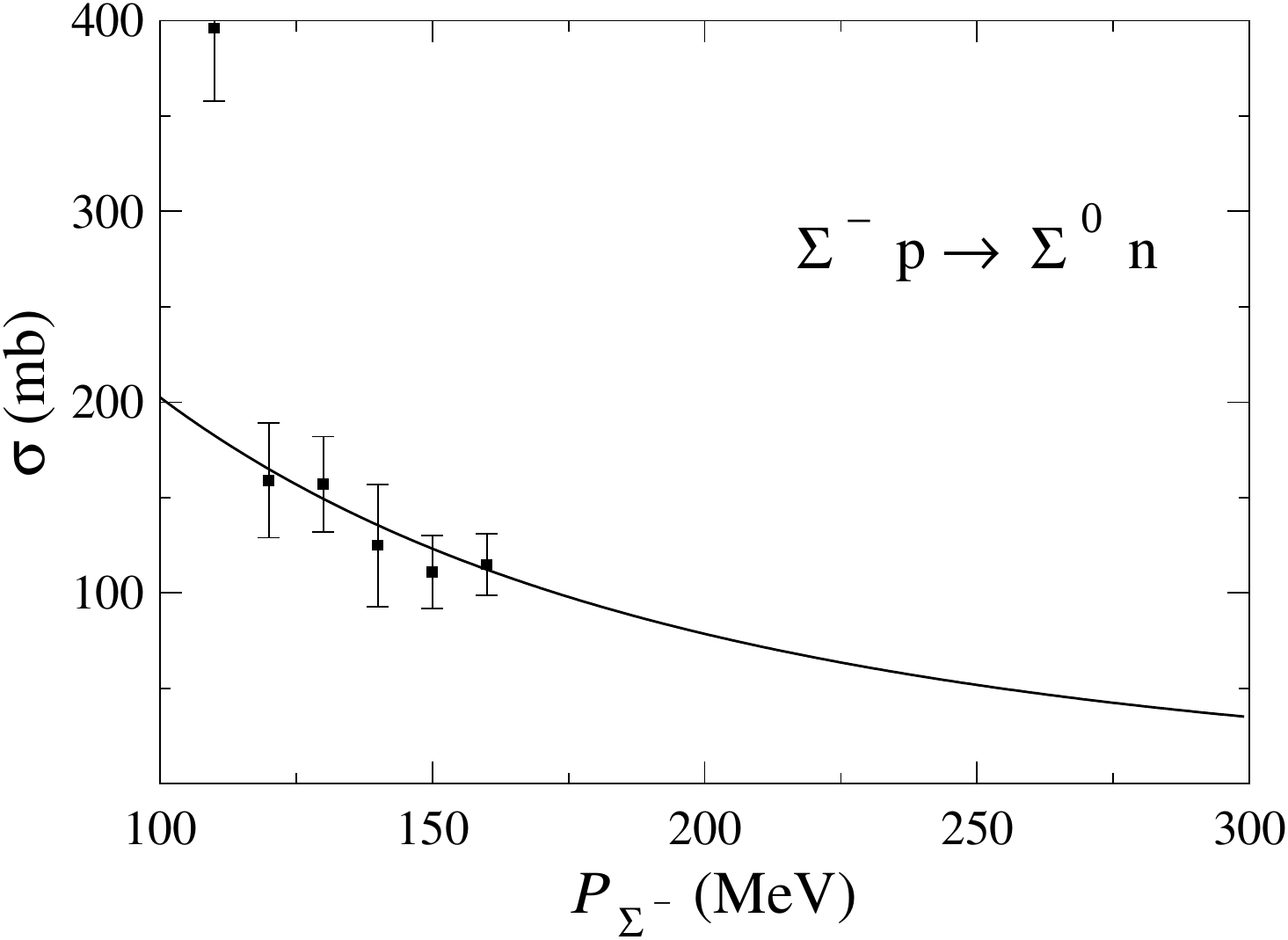} \\[3mm]
\includegraphics[width=0.40\textwidth]{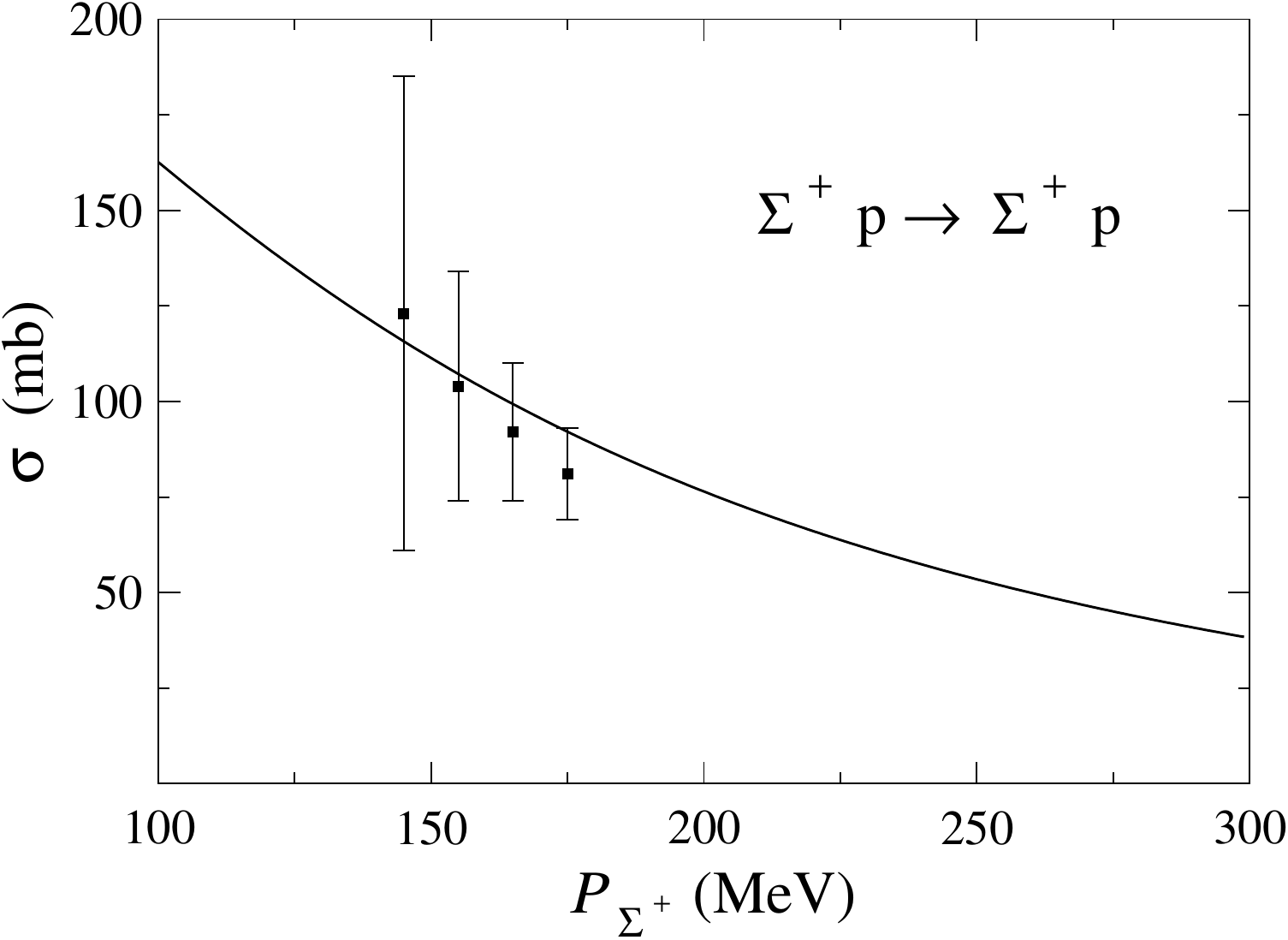}
 \end{center}

 \caption{Comparison of the theoretical $\Sigma N$ and $\Lambda N$ cross-sections given by 
our spin- and isospin-dependent potential  $V_{I,S}^{YN} = V_{\Sigma N-\Lambda N}^{\rm ScL,SDep}$
from Ref.\cite{FineTune_FBS}  (lines) with experimental
data~\cite{SigmaN1,SigmaN2,SigmaN3,SigmaN4,SigmaN5} (symbols).}
\label{YN_CrosSect.fig}
\end{figure*}

Since $\Sigma N$ system in isospin $I=1/2$ state is coupled to the $\Lambda N$ channel, 
the coupled-channel $\Sigma N - \Lambda N$ potential
\begin{equation}
\label{VSigN12}
 V_{I=1/2,S}^{\Sigma N-\Lambda N}(k,k') = g_{I,S}^{YN}(k) \, 
\Lambda_{I=1/2,S}^{\Sigma N - \Lambda N} \;  g_{I,S}^{YN}(k')
\end{equation}
was used. Here $\Lambda_{I=1/2,S}^{\Sigma N - \Lambda N}$ is a $2\times2$ matrix with elements
$\lambda_{I,S}^{\alpha_{YN} \beta_{YN}}$ and indices $\alpha_{YN},\beta_{YN} = 1 (\Sigma N)$
or $2 (\Lambda N)$. The potential $V_{I=1/2,S}^{\Sigma N - \Lambda N}$ is a $2\times 2$ matrix
as well.
The $\Sigma N$ interaction model in isospin $I=3/2$ state is a one-channel one
\begin{equation}
\label{VSigN32}
 V_{I=3/2,S}^{\Sigma N}(k,k') = \lambda_{I=3/2,S}^{YN} \;
g_{I,S}^{YN}(k) \, g_{I,S}^{YN}(k')
\end{equation}
with strength constants $\lambda_{I=3/2,S}^{\alpha_{YN}}$ and index $\alpha_{YN} = 1 (\Sigma N)$ only.
In both cases the Yamaguchi form factors
\begin{equation}
\label{VSigNFF}
g_{I,S}^{YN} (k)  =  \frac{1}{(k^2 + \beta_{I,S}^{YN})^2}
\end{equation}
were used.
The parameters of the spin-dependent 
$V_{\Sigma N-\Lambda N}^{\rm ScL,SDep}$ potential fitted to the $YN$ cross-sections and
scattering lengths can be found in Table~1 of~Ref.\cite{FineTune_FBS}. The 
$\Sigma N$ and $\Lambda N$ cross-sections compared with the experimental data are shown
in Fig. \ref{YN_CrosSect.fig} of the present paper. It is seen that the experimental data are reproduced
quite well. The scattering lengths given by our potential are presented in Table \ref{aYN.tab} together
with those from Ref.~\cite{HyperN_poten}. The scattering lengths given by
$V_{\Sigma N-\Lambda N}^{\rm ScL,SDep}$ are in agreement with the "advanced" ones
for $\Lambda N$ channel and $\Sigma N$ channel in spin $S=1$ state, while for the remaining 
channels the difference is quite large. Since values of the $YN$ scattering lengths are not measurable
values, which in addition are different in different models of advanced hyperon-nucleon potentials,
we do not consider it as a serious handicap of the model.
\begin{table}[ht]
\begin{center}
\begin{tabular}{ccc}
\hline  \noalign{\smallskip}
  &   $V_{I,S}^{YN}$  & "Advanced" $V^{YN}$ \cite{HyperN_poten} \\[1mm]
\noalign{\smallskip} \hline \noalign{\smallskip}
 $a_{I=1/2,S=0}^{\Sigma N}$ & $-1.40 + i \, 0.00$  & $-1.03 + i \, 0.00$ \\[2mm]
 $a_{I=1/2,S=1}^{\Sigma N}$ & $-0.03 + i \, 5.77$  & $-2.60 + i \, 2.56$  \\[2mm]
 $a_{I=3/2,S=0}^{\Sigma N}$ & $2.78$   & $3.47$ \\[2mm]
 $a_{I=3/2,S=1}^{\Sigma N}$   & $-0.37$  & $-0.41$ \\[2mm]
 $a_{I=1/2,S=0}^{\Lambda N}$ & $2.57$ & $$2.80 \\[2mm]
 $a_{I=1/2,S=1}^{\Lambda N}$ & $1.49$ & $1.56$ \\
 \noalign{\smallskip} \hline
\end{tabular}
\caption{Hyperon-nucleon scattering lengths (in fm) given by our spin- and isospin-dependent
potential  $V_{I,S}^{YN} = V_{\Sigma N-\Lambda N}^{\rm ScL,SDep}$ from Ref.\cite{FineTune_FBS} 
compared to those of the "advanced" potential \cite{HyperN_poten} (the sign rules in the papers 
are opposite).
\label{aYN.tab}
}
\end{center}
\end{table}

In our previous three-body calculations, only two particle channels $\bar{K} NN$
and $\pi \Sigma N$ were taken into account directly, 
due to this an exact optical version of the two-channel $V_{I=1/2,S}^{YN}$ potential was used.
The present studies were performed with three coupled three-body channels, including 
the $\pi \Lambda N$ one, it allowed us to use the hyperon-nucleon $I=1/2$ potential in its
original coupled-channel $\Sigma N - \Lambda N$ version.

\subsection{$\pi N$ potential}
\label{VpiN.sect}

Pion-nucleon $T$-matrix appears in the lower channels of the $\bar{K}NN$ systems, $\pi \Sigma N$ and
$\pi \Lambda N$. Since the pion-nucleon interaction in $s$-wave state is weak, it was neglected in our earlier
three-body calculations of the $K^- pp$ and $K^- np$ systems \cite{Kd_Kpp_last}.  Dependence of the three-body
characteristics of the quasi-bound state in the $K^- pp$ system was studied in our recent
paper~\cite{FineTune_FBS}, where several parametrizations of a one-term separable $\pi N$ potential
\begin{eqnarray}
\label{VpiN}
 V_{I}^{\pi N}(k,k') &=& \lambda_{I}^{\pi N} \;
g_{I}^{\pi N}(k) \, g_{I}^{\pi N}(k'), \\
g_{I}^{\pi N} (k) &=& \frac{1}{(k^2 + \beta_{I}^{\pi N})^2 }
\end{eqnarray}
were prepared and used. It turned out that the three-body characteristics depend on the pion-nucleon interaction
weakly. 

\begin{figure}
 \begin{center}
 \includegraphics[width=0.45\textwidth]{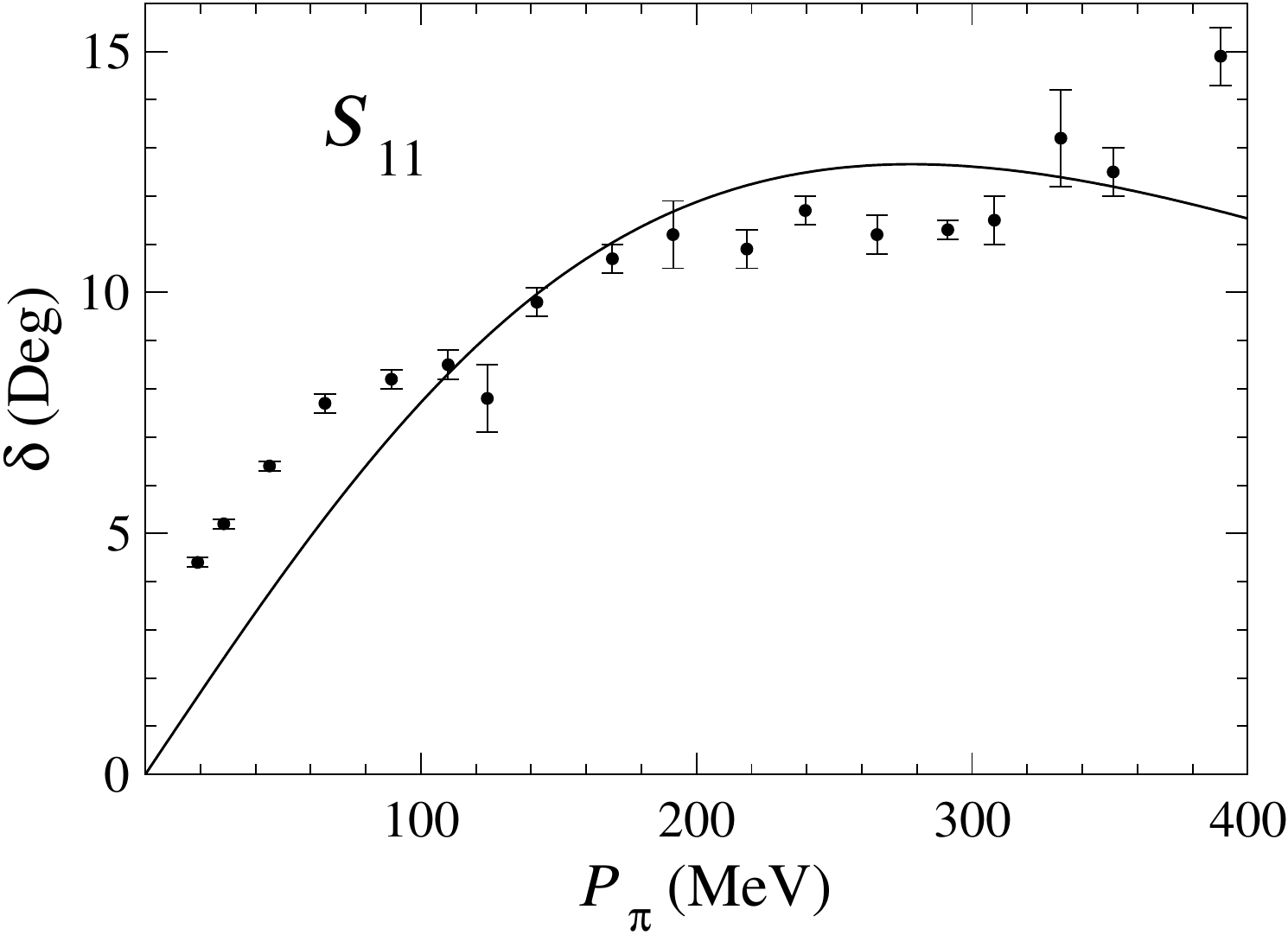}
\includegraphics[width=0.45\textwidth]{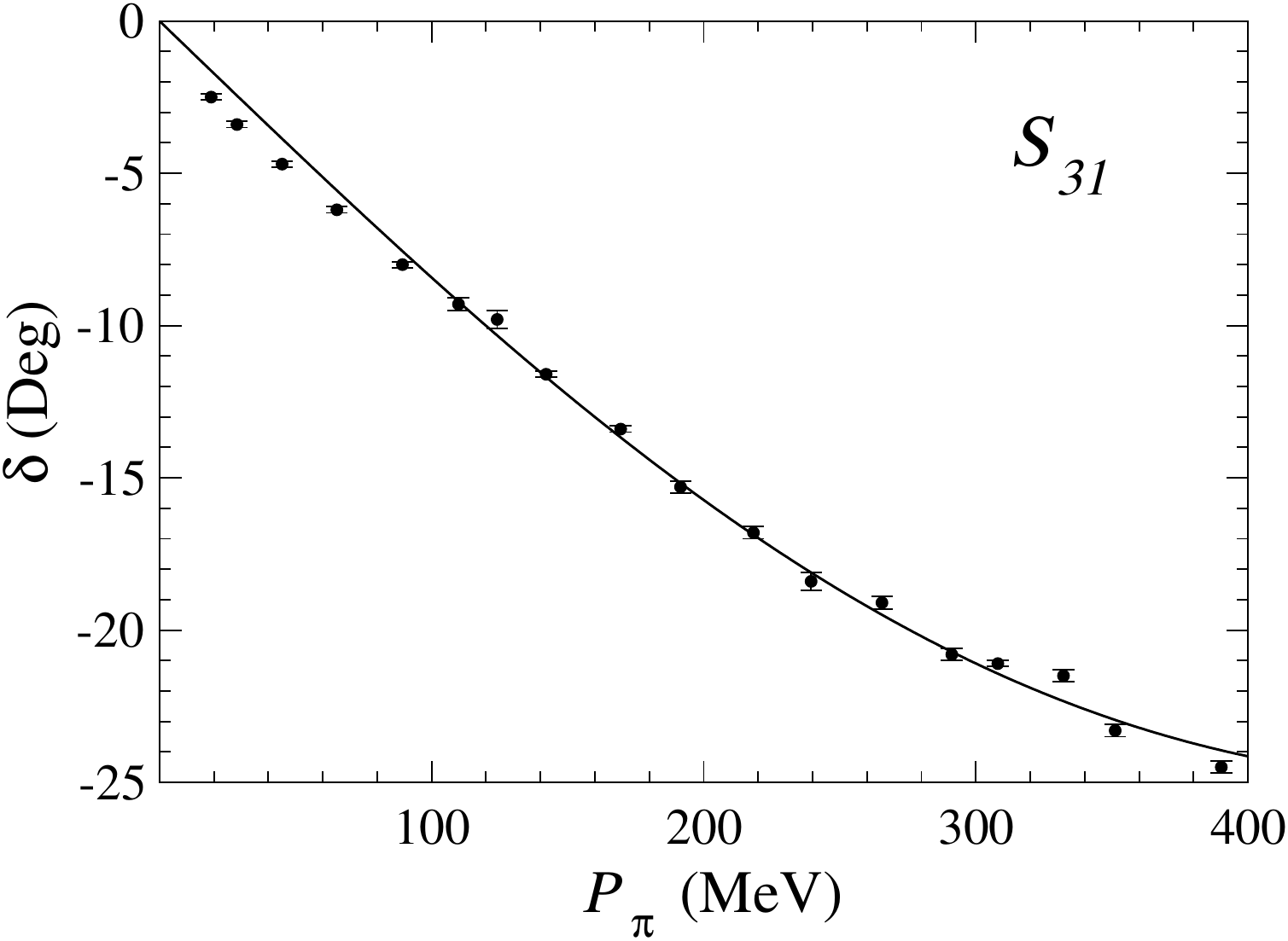}
 \end{center}

 \caption{Phase shifts of pion-nucleon scattering in $S_{11}$ and $S_{31}$ states:
given by our theoretical pion-nucleon potential Eq.(\ref{VpiN}) with parameters
Eq.(\ref{piN_params}) (lines) and existing ones~\cite{piN_phases2}(symbols).}
\label{piN_delta.fig}
\end{figure}

In the present calculations we used strength $\lambda_{I}^{\pi N}$ and range $\beta_{I}^{\pi N}$
parameters of the $V_{I}^{\pi N}$ potential fitted to existing data on $s$-wave phase shifts \cite{piN_phases1} 
corresponding to the WI08 fit of Ref. \cite{piN_phases2}
and to pion-nucleon scattering lengths \cite{piN_scL}. The parameters of the new $s$-wave pion-nucleon potential
Eq.(\ref{VpiN}) are
\begin{eqnarray}
\label{piN_params}
&{}& \beta_{1/2}^{\pi N} = 2.54 \, {\rm fm^{-1}},  \quad \lambda_{1/2}^{\pi N} =  -0.41 \, {\rm fm}, \\
&{}& \beta_{3/2}^{\pi N} = 2.74 \, {\rm fm^{-1}},  \quad \lambda_{3/2}^{\pi N} =  1.49 \, {\rm fm}.
\end{eqnarray}
The $s$-wave phase shifts given by our new $V_{I}^{\pi N}$ potential are in a good agreement with the WI08
fit~\cite{piN_phases2}, as can be seen in Fig. \ref{piN_delta.fig}.  The resulting scattering lengths in isospin $I=1/2$ and
$I=3/2$ states
\begin{equation}
a_{\pi N, I=1/2}^{\rm Th} = 0.34  \,{\rm fm}, \;
a_{\pi N, I=3/2}^{\rm Th} = -0.34 \, {\rm fm}
\end{equation}
reproduce the "experimental" values~\cite{piN_scL} 
\begin{equation}
a_{\pi N, I=1/2}^{\rm "Exp"} = 0.26 \, {\rm fm}, \;
a_{\pi N, I=3/2}^{\rm "Exp"} = -0.11 \, {\rm  fm}
\end{equation}
only approximately. Since the extraction of a scattering length from an experiment is not a trivial
task and can lead to quite inaccurate results (see e.g. \cite{ScLExtract} for the $K^- p$ case), we
are satisfied with this accuracy level. 

\section{Three-body results with three coupled particle channels and new two-body input}
\label{results_sect}

The results of the fine-tuned three-body calculations are presented in the
following subsections for the $K^- pp$, $K^-np$, and $\bar{K} \bar{K} N$ systems.
Binding energies and widths of the quasi-bound states in the three-body systems 
were obtained as solutions of Faddeev-type AGS equations with three coupled
particle channels and all necessary two-body interaction models described in
Section \ref{TwoBodyInput_sect} as an input. 
%
\begin{table*}
\begin{center}
\begin{tabular}{ccccccc}
\hline  \noalign{\smallskip}
  & \multicolumn{2}{c}{$V_{\bar{K}N}^{\rm 1,SIDD}$} 
    & \multicolumn{2}{c}{$V_{\bar{K}N}^{\rm 2,SIDD}$}  
     & \multicolumn{2}{c}{$V_{\bar{K}N}^{\rm Chiral}$}  
     \\[1mm]
  & $B_{K^-pp}$ \; &  $\Gamma_{K^-pp}$ \;  & $B_{K^-pp}$ \;  & $\Gamma_{K^-pp}$ \;  
        & $B_{K^-pp}$ \;  & $\Gamma_{K^-pp}$  \\[1mm]
\noalign{\smallskip} \hline \noalign{\smallskip}
   $V_{\rm Prev \Sigma N, \pi N}^{\rm 2ch}$ & $52.2$  & $67.1$ & $46.6$ & $51.2$ & $29.4$ & $46.4$   \\[2mm]
   $V_{\rm New \Sigma N, \pi N}^{\rm 3ch, A}$ & $44.3$  & $91.4$ & $43.4$ & $63.0$ & $24.7$ & $53.0$   \\[2mm]
   $V_{\rm New \Sigma N, \pi N}^{\rm 3ch, B}$ & $46.8$  & $91.4$ & $49.5$ & $68.6$ & $-$ & $-$   \\[1mm]
 \noalign{\smallskip} \hline
\end{tabular}
\caption{$K^- pp$ quasi-bound state: the new results of three-channel three-body 
$\bar{K}NN - \pi \Sigma N - \pi \Sigma \Lambda $ calculations 
(denoted as $V_{\rm New \Sigma N, \pi N}^{\rm 3ch}$)
using the new antikaon-nucleon potentials
$V_{\bar{K}N}^{\rm 1,SIDD}$, $V_{\bar{K}N}^{\rm 2,SIDD}$, $V_{\bar{K}N}^{\rm Chiral}$ with three coupled
channels together with the new $V_{\Sigma N-\Lambda N}$ and $V_{\pi N}$ potentials 
compared to the previous results ($V_{\rm Prev \Sigma N, \pi N}^{\rm 2ch}$) of two-channel three-body
$\bar{K}NN - \pi \Sigma N$ calculations \cite{Kd_Kpp_last} with older potentials. 
Binding energies $B_{K^-pp}$ (MeV) and widths $\Gamma_{K^- pp}$ (MeV) are shown.
A and B versions of the phenomenological one- $V_{\bar{K}N-\pi \Sigma - \pi \Lambda}^{\rm 1,SIDD}$ and 
two-pole $V_{\bar{K}N-\pi \Sigma - \pi \Lambda}^{\rm 2,SIDD}$ potentials are those with negative or posiive
$I=1$ strength constants, correspondingly.
}
\label{Kpp_res.tab}
\end{center}
\end{table*}

\subsection{Quasi-bound state in the $K^- pp$ system}
\label{Kpp_res.sect}

The main aim of the present study was to fine-tune the three-body calculations, particularly
the two-body input for them, in such a way that the three-body results better reproduce the E15
experimental data \cite{E15_JPARC2} on the binding energy and width of the 
$K^- pp$ quasi-bound state: $B_{K^- pp} = 42 \pm 3 \,^{+3}_{-4}$ MeV, 
$\Gamma_{K^- pp} = 100 \pm 7 \,^{+19}_{-9}$ MeV.

The three-body Faddeev-type AGS equations Eq.(\ref{AGS3ch}) with coupled $\bar{K}NN$, $\pi \Sigma N$,
and $\pi \Lambda N$ channels Eq.(\ref{KNNchnls}) in spin $S=0$ were solved. Three new models of
the antikaon-nucleon interaction with three coupled $\bar{K}N - \pi \Sigma - \pi \Lambda$ channels,
described in Section \ref{VKN} were used:
one- pole $V_{\bar{K}N - \pi \Sigma - \pi \Lambda}^{\rm 1,SIDD}$ and two-pole 
$V_{\bar{K}N - \pi \Sigma - \pi \Lambda}^{\rm 2,SIDD}$ phenomenological
potentials (A and B versions of them) and the chirally motivated potential
$V_{\bar{K}N - \pi \Sigma - \pi \Lambda}^{\rm Chiral}$.  
Two more interaction models entering the equations as the input are:
the hyperon-nucleon spin-dependent potential with parameters fitted to experimental $YN$
cross-sections and scattering lengths (Section \ref{VYN})
and the pion-nucleon $s$-wave potential describing $\pi N$ phase shifts and scattering lengths
(Section \ref{VpiN}). One more potential used as the input for the three-body AGS equations
is the TSN nucleon-nucleon potential \cite{Kd_Kpp_last} .

The results of our calculations are presented in Table \ref{Kpp_res.tab}. Our previous results obtained in the
two-channel three-body $\bar{K}NN - \pi \Sigma N$ calculations \cite{Kd_Kpp_last} with older antikaon-nucleon,
hyperon-nucleon potentials and switched off pion-nucleon interaction, are shown at the first line for comparison.
The results of the new three-body calculations with A and B versions of the new phenomenological
$\bar{K}N - \pi \Sigma - \pi \Lambda$  potentials are preseneted at lines two and three correspondingly.

It is seen that the new widths of the quasi-bound state in the $K^- pp$ system are much larger than
the previous ones for all new antikaon-nucleon potentials used in the calculations. 
Both: A and B versions of the new one-pole phenomenological antikaon-nucleon potential
(with negative/positive $I=1$ strength constants, see Section \ref{VKN} for details)
lead to the three-body width of the $K^- pp$ quasi-bound state, which reproduces the experimental
data from the E15 experiment at J-PARC \cite{E15_JPARC2}. 
Our attempts to find a set of parameters for the two-pole phenomenological 
antikaon-nucleon interaction model, which also reproduces the three-body experimental width together
with the two-body data, were unsuccessful. The three-body width from the calculations with
the two-pole phenomenological model is still much smaller than the experimental one.
The newly fitted chirally motivated $\bar{K}N - \pi \Sigma - \pi \Lambda$ potential lead to the width of
the $K^- pp$ quasi-bound state, which is larger than our previous results, but is almost twice smaller than
the experimental ones.

As for the binding energies, they are smaller than the previous ones, except those calculated with
B version of the two-pole phenomenological potential. The widths of the $K^- pp$ quasi-bound state
evaluated with the phenomenological models are quite close ones
to others, and they are in agreement with the experimental value from \cite{E15_JPARC2}.
The only exception is B version of the two-pole model of antikaon-nucleon interaction, which leads
to somewhat larger binding energy. The chirally motivated model gives a much smaller value than
the experimental one, which is smaller than the previous one.

Comparing the three-body results evaluated with A and B versions, which have negative and positive
$I=1$ strength constants correspondingly, we see that attraction or repulsion changes three-body
results only slightly when the one-pole version of the $\bar{K}N - \pi \Sigma - \pi \Lambda$ potential
is used. The two-pole phenomenological antikaon-nucleon potential leads to more visible differences for
three-body binding energies and widths. The negative isospin $I=1$ strength constants in the two-pole
potential lead to a smaller binding energy and width than the positive ones. Therefore, the attraction
causes a weaker bound and more stable quasi-bound $K^- pp$ quasi-bound state than the $I=1$ repulsion.

\subsection{Quasi-bound state in the $K^- pn$ system}
\label{Kpn_res.sect}

We reported in \cite{Kd_qbs} than another spin state $S=1$ of the $\bar{K}NN$ system, which can be denoted
as $K^- np$, also can have a quasi-bound state similar to that one in the $K^- pp$ system.
It is a state caused purely by strong interactions, in contrast to the atomic state of the $K^- np$ system,
a kaonic deuterium, which is mainly caused by the Coulomb interaction.
We solved the three-body AGS equations Eq.(\ref{AGS3ch}) with three coupled channels Eq.(\ref{KNNchnls})
using the new $\bar{K}N - \pi \Sigma - \pi \Lambda$,
$YN$ and $\pi N$ potentials together with $V^{TSN}_{NN}$. The results can be seen in Table \ref{Knp_res.tab}.
\begin{table*}
\begin{center}
\begin{tabular}{ccccccc}
\hline  \noalign{\smallskip}
  & \multicolumn{2}{c}{$V_{\bar{K}N}^{\rm 1,SIDD}$} 
    & \multicolumn{2}{c}{$V_{\bar{K}N}^{\rm 2,SIDD}$}  
     & \multicolumn{2}{c}{$V_{\bar{K}N}^{\rm Chiral}$}  
     \\[1mm]
  & $B_{K^-np}$ \; &  $\Gamma_{K^-np}$ \;  & $B_{K^-np}$ \;  & $\Gamma_{K^-np}$ \;  
        & $B_{K^-np}$ \;  & $\Gamma_{K^-np}$  \\[1mm]
\noalign{\smallskip} \hline \noalign{\smallskip}
   $V_{\rm Prev \Sigma N, \pi N}^{\rm 2ch}$ & $-$  & $-$ & $0.9$ & $59.4$ & $1.3$ & $41.8$   \\[2mm]
   $V_{\rm New \Sigma N, \pi N}^{\rm 3ch, A}$ & $6.0$  & $38.0$ & $6.0$ & $32.4$ & $2.7$ & $35.6$   \\[2mm]
   $V_{\rm New \Sigma N, \pi N}^{\rm 3ch, B}$ & $6.1$  & $38.0$ & $6.0$ & $34.0$ & $-$ & $-$   \\[1mm]
 \noalign{\smallskip} \hline
\end{tabular}
\caption{$K^- np$ quasi-bound state:  the new results of three-channel three-body 
$\bar{K}NN - \pi \Sigma N - \pi \Sigma \Lambda $ calculations 
(denoted as $V_{\rm New \Sigma N, \pi N}^{\rm 3ch}$)
using the new antikaon-nucleon potentials
$V_{\bar{K}N}^{\rm 1,SIDD}$, $V_{\bar{K}N}^{\rm 2,SIDD}$, $V_{\bar{K}N}^{\rm Chiral}$ with three coupled
channels together with the new $V_{\Sigma N-\Lambda N}$ and $V_{\pi N}$ potentials 
compared to the previous results ($V_{\rm Prev \Sigma N, \pi N}^{\rm 2ch}$) of two-channel three-body
$\bar{K}NN - \pi \Sigma N$ calculations \cite{Kd_Kpp_last} with older potentials. 
Binding energies $B_{K^-np}$ (MeV) and widths $\Gamma_{K^- np}$ (MeV) are shown.
A and B versions of the phenomenological one- $V_{\bar{K}N-\pi \Sigma - \pi \Lambda}^{\rm 1,SIDD}$ and 
two-pole $V_{\bar{K}N-\pi \Sigma - \pi \Lambda}^{\rm 2,SIDD}$ potentials are those with negative or posiive
$I=1$ strength constants, correspondingly.
}
\label{Knp_res.tab}
\end{center}
\end{table*}

In contrast to our previous calculations \cite{Kd_Kpp_last}, when the one-pole phenomenological antikaon-nucleon
potential did not form a quasi-bound state in the $K^-np$ system, all antikaon-nucleon potentials now lead to
the existence of  the quasi-bound state. The new widths of the $K^- np$ quasi-bound state provided by 
antikaon-nucleon models with two poles corresponding to the $\Lambda(1405)$ resonance 
($V_{\bar{K}N}^{\rm 2,SIDD}$ and $V_{\bar{K}N}^{\rm Chiral}$) are much smaller than the previously predicted
ones. Now the three-body binding energies are  quite close to each other for all antikaon-nucleon interacion
models. The binding energies calculated with the two-pole models are much larger than the previous ones.
They are almost identical for the results obtained with the phenomenological one- and two-pole
$\bar{K}N - \pi \Sigma - \pi \Lambda$ interaction models, while the chirally motivated potential leads to much
smaller binding energy than the phenomenological ones.
The differences between the results of calculations with A and B versions of the phenomenological potentials
are very small for the $K^- np$ system not only for the binding energies, as in the $K^- pp$ system,
but for the widths as well.

\subsection{Quasi-bound state in the $\bar{K} \bar{K} N$ system}
\label{KKN_res.sect}

One more three-body system studied by us previously \cite{KKN} is $\bar{K} \bar{K} N$ system with double
strangeness,  which is $K^- K^- p$ in particle representation. In this case the Faddeev-type AGS equations Eq.(\ref{AGS3ch})
with coupled three-body channels $\bar{K} \bar{K} N$, $\bar{K} \pi \Sigma$, and $\bar{K} \pi \Lambda$
Eq.(\ref{KKNchnls}) were solved. The new models of antikaon-nucleon interaction directly coupling three
particle channels were used together with the previously constructed and used $\bar{K} \bar{K}$ potential.
The remaining interactions in the lowest three-body $\bar{K} \pi \Sigma$ and $\bar{K} \pi \Lambda$ channels, 
which are $\bar{K} \pi$ and $\bar{K} Y$  ones, are not known. We assumed that they are not so important
and, as previously, switched them off.
\begin{table*}
\begin{center}
\begin{tabular}{ccccccc}
\hline  \noalign{\smallskip}
  & \multicolumn{2}{c}{$V_{\bar{K}N}^{\rm 1,SIDD}$} 
    & \multicolumn{2}{c}{$V_{\bar{K}N}^{\rm 2,SIDD}$}  
     & \multicolumn{2}{c}{$V_{\bar{K}N}^{\rm Chiral}$}  
     \\[1mm]
  & $B_{\bar{K}\bar{K}N}$ \; &  $\Gamma_{\bar{K}\bar{K}N}$ \;  & $B_{\bar{K}\bar{K}N}$ \;  & $\Gamma_{\bar{K}\bar{K}N}$ \;  
        & $B_{\bar{K}\bar{K}N}$ \;  & $\Gamma_{\bar{K}\bar{K}N}$  \\[1mm]
\noalign{\smallskip} \hline \noalign{\smallskip}
   $V^{\rm 2ch}$ & $19.51$  & $102.02$ & $25.93$ & $84.60$ & $16.09$ & $61.32$   \\[2mm]
   $V^{\rm 3ch, A}$ & $5.5$  & $80.0$ & $14.5$ & $82.0$ & $9.4$ & $57.4$   \\[2mm]
   $V^{\rm 3ch, B}$ & $9.4$  & $88.4$ & $20.0$ & $87.6$ & $-$ & $-$   \\[1mm]
 \noalign{\smallskip} \hline
\end{tabular}
\caption{$\bar{K}\bar{K}N$ quasi-bound state:  the new results of three-channel three-body 
$\bar{K}\bar{K}N - \bar{K}\pi \Sigma - \bar{K} \pi \Lambda $ calculations 
(denoted as $V^{\rm 3ch}$) using the new antikaon-nucleon potentials
$V_{\bar{K}N}^{\rm 1,SIDD}$, $V_{\bar{K}N}^{\rm 2,SIDD}$, $V_{\bar{K}N}^{\rm Chiral}$ with three coupled
channels 
compared to the previous results ($V^{\rm 2ch}$) of two-channel three-body
$\bar{K}\bar{K}N - \bar{K} \pi \Sigma$ calculations with older potentials \cite{KKN}. 
Binding energies $B_{\bar{K}\bar{K}N}$ (MeV) and widths $\Gamma_{\bar{K}\bar{K}N}$ (MeV) are shown.
A and B versions of the phenomenological one- $V_{\bar{K}N-\pi \Sigma - \pi \Lambda}^{\rm 1,SIDD}$ and 
two-pole $V_{\bar{K}N-\pi \Sigma - \pi \Lambda}^{\rm 2,SIDD}$ potentials are those with negative or posiive
$I=1$ strength constants, correspondingly.
}
\label{KKN_res.tab}
\end{center}
\end{table*}

The new three-body results for the binding energy $B_{\bar{K}\bar{K}N}$ and width $\Gamma_{\bar{K}\bar{K}N}$ 
of the quasi-bound state in the $\bar{K} \bar{K}N$ system are presented in Table \ref{KKN_res.tab} together with
our previous results  from Ref.~\cite{KKN}. It is seen that in the $K^- K^- p$ system, in contrast to the $K^- pp$ one,
the new three-body equations with new antikaon-nucleon potentials lead to smaller both: binding energies and widths,
for almost all antikaon-nucleon potentials used as the input (except the width calculated with B version of 
$V_{\bar{K}N}^{\rm 2,SIDD}$). It is true especially for the results evaluated with the
one-pole phenomenological $\bar{K}N - \pi \Sigma - \pi \Lambda$ potential, where the differences are large.

The resulting widths obtained with the phenomenological potentials are quite close to each other for the A and
B versions. Similar to the $K^- pp$ quasi-bound state, the three-body $K^- K^- p$ widths calculated with
A versions are smaller than those with B versions. The corresponding binding energies are also smaller when
a phenomenological model with negative $I=1$ strength constants is used, comparing to those with the positive
ones.

The chirally motivated model of antkaon-nucleon interaction give a smaller width than the phenomenological ones
As for the binding energy, that one of the $\bar{K} \bar{K} N$ quasi-bound state given by the chirally motivated
antikaon-nucleon potential is not the smallest one, as it is in the $K^- pp$ and 
$K^- np$ systems.

Our previous results for the double antikaon nucleon system \cite{ KKN} were roughly consistent with
parameters of a quasibound state found in an experiment. The newly evaluated total masses of the
$\bar{K}\bar{K}N$ system and their widths are consistent with another experimental result
\cite{Ksi1950_exp}. The experiment was devoted to the isospin one half $\Xi(1950)$ resonance (denoted
as $\Xi(1940)$ in the paper) studied in the $K^- p$ reaction. Our new masses predicted by all three
antikaon-nucleon potentials, except the $V_{\bar{K}N}^{\rm 2,SIDD-B}$ version, are within the experimental
region of the measured mass $M_{\Xi(1950)} = 1936 \pm 22$ MeV (the threshold energy for the
$\bar{K}\bar{K}N$ system is $1930.2$ MeV). In addition, the widths obtained with
the phenomenological potentials are within the experimental region of the experimental width
$\Gamma_{\Xi(1950)} = 87 \pm 26$ MeV.

\subsection{Joint three-body results}
\label{Joint3body_res.sect}

The obtained three-body results for the $K^- pp$, $K^- np$ and $K^- K^- p$ systems from Tables \ref{Kpp_res.tab},
 \ref{Knp_res.tab}, and \ref{KKN_res.tab} are presented all together in  Table \ref{2and3bodyPoles_res.tab} for
convinience. Looking at, it we can see that the $K^- pp$ is the system which has the largest binding energies
independently of the antikaon-nucleon interaction used in the calculations. On the contrary, the $K^- np$ system
has the smallest width.
\begin{table*}[ht]
\begin{center}
\begin{tabular}{ccccccccccc}
\hline  \noalign{\smallskip}
  & \multicolumn{2}{c}{$V_{\bar{K}N}^{\rm 1,SIDD-A}$} 
  & \multicolumn{2}{c}{$V_{\bar{K}N}^{\rm 1,SIDD-B}$} 
    & \multicolumn{2}{c}{$V_{\bar{K}N}^{\rm 2,SIDD-A}$} 
    & \multicolumn{2}{c}{$V_{\bar{K}N}^{\rm 2,SIDD-B}$}  
     & \multicolumn{2}{c}{$V_{\bar{K}N}^{\rm Chiral}$}  
     \\[1mm]
  & $\Delta E_{1s}$ \; &  $\Gamma_{1s}$ \;  & $\Delta E_{1s}$ \;  & $\Gamma_{1s}$ \;  
& $\Delta E_{1s}$ \; &  $\Gamma_{1s}$ \;  & $\Delta E_{1s}$ \;  & $\Gamma_{1s}$ \;  
        & $\Delta E_{1s}$ \;  & $\Gamma_{1s}$  \\[1mm]
\noalign{\smallskip} \hline \noalign{\smallskip}
   $(K^- p)_{Coul}$ &  $-314.1$ \,  &  $626.9$  &  $-306.4$  \,   &  $600.7$  &
 $-293.8$  \, &  $639.8$  &  $-292.5$ \,  &  $609.8$   &  $-319.9$ \,  &  $622.0$   \\[1mm]
 \noalign{\smallskip} \hline
  & $z_{tot}$ \; &  $\Gamma$ \;  & $z_{tot}$ \;  & $\Gamma$ \;  
  & $z_{tot}$ \; &  $\Gamma$ \;  & $z_{tot}$ \;  & $\Gamma$ \;  
        & $z_{tot}$ \;  & $\Gamma$  \\[1mm]
\noalign{\smallskip} \hline \noalign{\smallskip}
   $(K^- p)_1$ & $1429.2$  & $67.0$ & $1427.6$  & $73.8$ & $1428.2$ & $81.8$ & 
$1426.8$  & $91.6$ & $1424.5$ & $58.0$   \\[2mm]
   $(K^- p)_2$ & $-$  & $-$ & $-$  & $-$ & $1355.3$ & $191.2$ & $1353.1$ & $175.6$ & $1376.1$ & $168.2$   \\[1mm]
 \noalign{\smallskip} \hline
  & $B$ \; &  $\Gamma$ \;  & $B$ \;  & $\Gamma$ \;  
 & $B$ \; &  $\Gamma$ \;  & $B$ \;  & $\Gamma$ \;  
        & $B$ \;  & $\Gamma$  \\[1mm]
\noalign{\smallskip} \hline \noalign{\smallskip}
   $K^- pp$ & $44.3$ & $91.4$ & $46.8$ & $91.4$ & $43.4$ & $63.0$ & $49.5$ & $68.6$ & $24.7$ & $53.0$   \\[2mm]
   $K^- np$ & $6.0$  & $38.0$ & $6.1$  & $38.0$ & $6.0$ & $32.4$ & $6.0$  & $34.0$ & $2.7$ & $35.6$   \\[2mm]
   $K^- K^- p$ & $5.5$  & $80.0$ & $9.4$  & $88.4$ & $14.5$ & $82.0$ & $20.0$  & $87.6$ & $9.4$ & $57.4$   \\[1mm]
 \noalign{\smallskip} \hline
\end{tabular}
\caption{Two-body kaonic hydrogen $1s$ level shift $\Delta E_{1s}$ (eV) and width $\Gamma_{1s}$ (eV), and
strong pole characteristics $z_{tot}$ (MeV) and  $\Gamma$ (MeV) corresponding to the $\Lambda(1405)$ resonance
denoted as $(K^- p)_i$ with $i=1,2$ together with the three-body binding energies $B$ (MeV) and widths
$\Gamma$ (MeV) of $\bar{K}NN$ and $\bar{K}\bar{K}N$ systems.
}
\label{2and3bodyPoles_res.tab}
\end{center}
\end{table*}

Binding energies of the $K^- K^- p$ system are larger or comparable with those of $K^- np$ system.
Th widths of the $K^- pp$ and $K^- K^- p$ depend of the number of poles of the $\bar{K}N - \pi \Sigma - \pi \Lambda$
potential. When the one-pole phenomenological $V_{\bar{K}N}^{\rm 1,SIDD}$ potential is used, the width
of the $K^- pp$ system is larger or comparable with that of the $K^- K^- p$ system. The situation is opposite
for the two-pole phenomenological $V_{\bar{K}N}^{\rm 2,SIDD}$ and the chirailly motivated $V_{\bar{K}N}^{\rm Chiral}$
potentials: the width of the system with double strangeness $K^- K^- p$ system is larger or comparable with that
one of the $K^- pp$ system.

The chirally motivated antikaon-nucleon potential leads to the smallest binding energy for both $\bar{K}NN$ 
systems and to the smallest width for the $K^- pp$ and $K^- K^- p$ systems. The A versions of the phenomenological
one- and two-pole potentials, which have negative isospin $I=1$ strength constants, give smaller or comparable 
binding energies and widths for all three three-body systems.

The negative isospin one strength constants in both antikaon-nucleon phenomenological potentials,
meaning the attractive $I=1$ interactions, give smaller or comparable binding energies and widths for all three
three-body systems that have the positive (repulsive) contants.

On the whole, the $K^- np$ is the least sensitive to the  $\bar{K}N - \pi \Sigma - \pi \Lambda$ interaction
model system, while the $K^- pp$ is the opposite. Probably, it is caused by the fact that the quasi-bound state pole for
the first system is situated much closer to the threshold than the remaining two, while that of the $K^- pp$
is the furthest.

Table \ref{2and3bodyPoles_res.tab} also contains two-body kaonic hydrogen $1s$ level shifts $\Delta E_{1s}$ and
widths $\Gamma_{1s}$, and strong pole characteristics corresponding to the $\Lambda(1405)$ resonance
of our antikaon-nucleon potentials used as the input for the three-body calculations.
Comparing the three-body binding energies and widths for the $K^- pp$, $K^- np$, and $K^- K^- p$ systems
with the two-body poles of the new $\bar{K}N - \pi \Sigma - \pi \Lambda$ potentials, we see no 
correlations between them. A naive conseption of the $K^- pp$ system as a ''doubled $\Lambda(1405)$''
resonance is far from reality. The three-body dynamics together with the coupled particle channels plays
an important role in all three three-body systems.

\section{Conclusions}
\label{Conclusions_sect}

The fine-tuning of the three-body Faddeev-type calculations of the quasi-bound state in the $K^- pp$ system
and the two-body input for them was successful. The binding energy and width of the $K^- pp$ quasi-bound
state evaluated with our new phenomenological potential with one-pole structure of the $\Lambda(1405)$
resonance  $V_{\bar{K}N-\pi \Sigma - \pi \Lambda}^{\rm 1,SIDD}$ reproduce the experimental data by E15
experiment from JPARC \cite{E15_JPARC2}. The other two new antikaon-nucleon potentials
$V_{\bar{K}N}^{\rm 2,SIDD}$ and $V_{\bar{K}N}^{\rm Chiral}$ give smaller widths.
The three-body Faddeev-type AGS equations with three coupled channels were solved with new models of
the antikaon-nucleon, $YN$, and $\pi N$ interactions, and previously used nucleon-nucleon or antikaon-antikaon
potentials.

We recalculated binding energies and width of the quasi-bound states in two other three-body systems: $K^- np$
and $K^- K^- p$ ones. In contrast to our previous results, all new antikaon-nucleon potentials now led to
the existence of the quasi-bound state in the spin one $\bar{K}NN$ system. The $K^- np$ system has the smallest
widths among the three three-body systems, therefore being the most stable one. The $K^- pp$ system
is characterized by the strongest binding energy among the three. The quasi-bound state in the $K^- K^- p$ could
be connected with one of the $\Xi$ states.

We also showed that the negative (attractive) isospin one strength constants of the phenomenological
$\bar{K}N - \pi \Sigma - \pi \Lambda$ potential lead to smaller or comparable binding energy and width of 
the quasi-bound states in $K^- pp$, $K^- np$, and $K^- K^- p$ systems comparing to the model
with the positive (repulsive) constants.


\newpage

\end{document}